\documentclass[aps,prb,twocolumn,showpacs,groupedaddress]{revtex4}
\usepackage[dvips]{graphicx}
\usepackage{amsmath}

\renewcommand{\Im}{\text{Im}}
\begin{document}
\title{Self-energy approach to the correlated Kondo-lattice model}
\author{W. Nolting$^{a}$,\ G. G. Reddy$^{b}$,\ A. Ramakanth$^{b}$,\
  D. Meyer$^{c}$,\ and J. Kienert$^{a}$}
\affiliation{$^{a}$\mbox{Lehrstuhl Festk{\"o}rpertheorie, Institut f{\"u}r
  Physik, Humboldt-Universit{\"a}t zu Berlin, Invalidenstr. 110, 110115 Berlin, Germany}\\
 $^{b}$\mbox{Kakatiya University, Department of Physics, Warangal-506009, India}\\
 $^{c}$\mbox{Department of Mathematics, Imperial College,\\ 180 Queen's Gate, London SW7 2BZ, United Kingdom}}
\date{\today}
\begin{abstract}
We develop an interpolating self-energy approach to the correlated
Kondo-lattice model. The correlation of the band electrons is taken into
account by a Hubbard interaction. The method is based on a self-energy
ansatz, the structure of which allows to fulfill a maximum number of
exactly solvable limiting cases. The parameters of the ansatz are fitted
to spectral moments via high-energy expansion of the self-energy. The
band electron correlations are taken into account by an effective medium
approach being correct in the strong coupling ($U$) regime. The theory
is considered reliable for all temperatures, band occupations, and
exchange couplings. Results are presented for the respective
dependencies of spectral densities, quasiparticle densities of states,
and characteristic correlation functions, and interpreted in terms of
elementary spin exchange processes between itinerant conduction
electrons and localized magnetic moments. The appearance of magnetic
polarons, the typical quasiparticle of Kondo-lattices, in the energy
spectrum is worked out. Spin exchange processes prevent a total spin
polarization of the band electrons even for arbitrarily strong exchange
couplings as long as the local moments are represented by quantum
mechanical spins. 
\end{abstract}
\pacs{75.10.-b, 71.10.Fd, 75.30.Et, 75.30.Mb}
\maketitle
\section{Introduction}
\label{sec1}
The Kondo-lattice model
(KLM)\cite{ZEN51a,ANH55,KAS56,KON64,RHB67,NOL79,DDN98} is surely
one of the most intensively 
discussed models in solid state theory. It aims at the mutual influence
of two electronic subsystems. The one consists of itinerant particles in
a partially filled energy band, the other is built up by at certain
lattice sites strictly localized electrons. The latter give rise to
permanent magnetic moments which are described as quantum mechanical
spins. Characteristic model properties result from an interband
exchange, which is written as an intra-atomic interaction of strength
$J$ between the conduction electron spin
\boldmath$\sigma$\unboldmath$_{i}$ and local-moment spin
$\mathbf{S}_{i}$: 

\begin{equation}
\label{eq:Hsf}
H_{sf}=-J\sum_{i}\mbox{\boldmath${\sigma}$}_{i}\cdot \mathbf{S}_{i}.
\end{equation}

The index $i$ refers to the lattice site $\mathbf{R}_{i}$. According to
the sign of the exchange coupling $J$, a parallel ($J$$>$0) or an
antiparallel ($J$$<$0) alignment of itinerant and 
localized spin is favoured with remarkable differences in the physical
properties. We restrict our considerations in the following to the  
$J$$>$0-case, sometimes referred to as \textit{ferromagnetic Kondo-lattice
  model} ($s$-$f$, $s$-$d$ model), or, in the strong-coupling regime,
as \textit{double-exchange model}.

The interest in the KLM explains itself by rather manifold
applications. There are mainly four groups of materials, the fundamental
properties of which are considered to be well-described by the exchange
(\ref{eq:Hsf}):
\begin{enumerate}
\renewcommand{\labelenumi}{(\alph{enumi})}
\item (Concentrated) magnetic semiconductors EuO, EuS, ..
\item Local-moment metals Gd, Tb, Dy, Eu$_{1-x}$Gd$_{x}$S, ..
\item Diluted magnetic semiconductors (DMS) Ga$_{1-x}$Mn$_{x}$As, ..
\item Manganites (colossal magneto-resistance materials, CMR)
  La$_{1-x}$(Ca,Sr)$_{x}$MnO$_{3}$ 
\end{enumerate}

The rich variety of hot topics in solid state theory, being related to
the KLM, deserves a reliable approach to this not exactly solvable
many-body problem. Most of recent theoretical works on the KLM, aiming 
predominantly at the CMR-materials, assume classical spins 
($S$$\rightarrow$$\infty$).\cite{FUR99,MLS95,HEV00} However, this  
seems to be a rather questionable starting point, only
\textit{justified} by the fact that then \textit{dynamical mean field
  theory} (DMFT) can be applied. Characteristic features such as magnon
emission (absorption) of the conduction electron are ruled out from the
very beginning. The same holds for the formation of \textit{magnetic
  polarons}, the characteristic quasiparticles of Kondo-lattices. The
importance of such effects can impressively be demonstrated by  
the exactly solvable limiting case of a single electron in an otherwise
empty conduction band, interacting with a ferromagnetically saturated
spin system \cite{SM81,NDM85} \mbox{(see Figs.~1,~2 in
Ref.~\onlinecite{MSN01}).} Recently we proposed a DMFT-based approach to
the KLM with quantum spins using a fermionization of the local spin
operators ($S$=1/2). Even for rather moderate exchange coupling an
unusual quasiparticle structure appears due to the above-mentioned
elementary processes. The results are in remarkable agreement with those
from the \textit{moment conserving decoupling approach} (MCDA),
introduced in Ref.~\onlinecite{NRM97} and successfully applied to  
several topics in previous
papers.\cite{ScN01,MN02,REN99,SN02,SCN01a,SMN01} 
The MCDA as well as the DMFT-based approach are both considered to work
in the low and intermediate coupling regime,\cite{MSN01} therefore
suitable for modelling systems such as EuO, EuS,
Gd.\cite{ScN01,MN02,REN99,SCN01a} For strongly coupled systems (DMS,
CMR-materials) with finite band occupations alternatives are
desirable. Very recently the authors have presented a self-energy
approach which interpolates between a maximum number of exactly solvable
limiting cases.\cite{NRRM01} On the one hand, it should be a reasonable
ansatz for weak as well as strong couplings, on the other hand, however,
it represents a low-density approach, strictly justified for
$n$$\rightarrow$0 only. We are going to present in this paper a new and
extended ansatz, which is considered to work for all couplings and all
band occupations reasonably well. For $n$$\rightarrow$0 it turns out to be
identical to the ansatz in Ref.~\onlinecite{NRRM01}.

The inclusion of finite band occupations provokes some non-trivial
problems. First, some physically important correlation functions
appear, which have to be determined self-consistently via spectral
theorem and properly defined \textit{higher} Green functions. These are,
in particular, \textit{mixed} correlations, i.e. averages of products of
local spin operators with band electron operators. More fundamental,
however, is another point: In the \textit{normal} KLM the band electrons
are considered as uncorrelated particles. It can be shown, directly to
be seen in the exact zero-bandwidth limit,\cite{NOM84} that the
transition from a single- to a double-occupied site can unphysically be
connected with a gain in energy. In reality, the Coulomb repulsion will
rather block such double occupancies. To guarantee this in the model,
too, we shall add to the \textit{normal} KLM an intra-atomic Coulomb
interaction as has been proposed in Ref.~\onlinecite{HEV00}. One then
speaks of the \textit{correlated Kondo-lattice model} (CKLM). 

In Sec.~\ref{sec2} we first formulate in detail the Hamiltonian of the 
CKLM and its many-body problem. Furthermore, we list up some rigorous
facts which are used in Sec.~\ref{sec3} to create an interpolation
formula for the electronic self-energy. This interpolation scheme is the
crucial point of our procedure. The resulting data are finally discussed
in Sec.~\ref{sec4}, mainly in terms of spectral densities (SD) and
quasiparticle densities of states (QDOS). The above-mentioned rigorous
facts have partly been presented already in our preceding
paper.\cite{NRRM01} So we restrict the presentation to the points which
are vital for the understanding of the following theory and the
subsequent discussions. 

\section{Theory}
\label{sec2}
\subsection{The many-body problem}
\label{subsec1}
The Hamiltonian of the \textit{correlated Kondo-lattice model} (CKLM)  
consists of three partial operators:  
\begin{equation}
\label{eq:H}
H = H_{S} + H_{U} + H_{sf}.
\end{equation}
$H_{S}$ is the kinetic energy of itinerant band electrons:
\begin{equation}
\label{eq:Hs}
H_{S} =
\sum_{ij\sigma}(T_{ij}-\mu\delta_{ij})c_{i\sigma}^{\dagger}c_{j\sigma}=\sum_{{\bf
    k}\sigma}(\epsilon({\bf k})-\mu)c_{{\bf k}\sigma}^{\dagger}c_{{\bf k}\sigma}^{}. 
\end{equation}
$T_{ij}$ are the hopping integrals being connected by Fourier
transformation to the \textit{free} Bloch energies 
$\epsilon(\mathbf{k})$. $\mu$ is the chemical
potential. $c^{\dagger}_{i\sigma}$ ($c_{i\sigma}$) and
$c^{\dagger}_{{\bf k}\sigma}$ ($c_{{\bf k}\sigma}$)   
are the creation (annihilation) operators of a conduction electron with
spin $\sigma$ ($\sigma=\uparrow, \downarrow$) at lattice site
$\mathbf{R}_{i}$ and with wavevector ${\bf k}$, respectively.  

$H_{U}$ is an intra-atomic Coulomb interaction of Hubbard-type with 
the Coulomb-matrix element $U$: 
\begin{equation}
\label{eq:Hu}
H_{U} = \frac{U}{2}\sum_{i\sigma}n_{i\sigma}n_{i-\sigma}.
\end{equation}
$n_{i\sigma} = c_{i\sigma}^{\dagger}c_{i\sigma}$ is the occupation
number operator. 

Most important is $H_{sf}$, already introduced by
(\ref{eq:Hsf}). However, better tractable is the second quantized form
of the exchange interaction: 
\begin{equation}
\label{eq:Hsf2ndquant}
H_{sf} = -\frac{1}{2}J\sum_{j\sigma}(z_{\sigma}S_{j}^{z}n_{j\sigma} +
S_{j}^{-\sigma}c_{j-\sigma}^{\dagger}c_{j\sigma}).
\end{equation}
Here we have written for abbreviation:
\begin{equation}
\label{eq:abbrev}
z_{\sigma} = \delta_{\sigma\uparrow} - \delta_{\sigma\downarrow}~;~  
S_{j}^{\sigma} = S_{j}^{x} + iz_{\sigma}S_{j}^{y}.
\end{equation} 
The first term in~(\ref{eq:Hsf2ndquant}) describes an Ising-like
interaction between the $z$-components of the localized and the 
itinerant spins. The second term is responsible for spin exchange
processes between the two subsystems.  
 
Since we are mainly interested in the conduction electron properties, we
try to calculate the single electron Green function, from which we can
derive all required information. A formal solution for the
${\bf k}$-dependent single-electron Green function can be written as 
follows:   
\begin{equation}
\label{eq:GformalS}
G_{\mathbf{k}\sigma}(E)=\left<\left<c_{\mathbf{k}\sigma};c_{\mathbf{k}\sigma}^{\dagger}\right>\right>_{E}
=\frac{\hbar}{E+\mu-\epsilon(\mathbf{k})-\Sigma_{\mathbf{k}\sigma}(E)}.
\end{equation}
The self-energy $\Sigma_{\mathbf{k}\sigma}(E)$ is composed of two parts:
\begin{equation}
\label{eq:selfE}
\Sigma_{\mathbf{k}\sigma}(E) = \Sigma_{\mathbf{k}\sigma}^{(U)}(E) +
\Sigma_{\mathbf{k}\sigma}^{(sf)}(E). 
\end{equation}
The two self-energy parts are defined by the following relation:
\begin{equation}
\label{eq:selfEparts}
\left<\left<\left[H_{U(sf)}, c_{\mathbf{k}\sigma}\right]_{-};
    c_{\mathbf{k}\sigma}^{\dagger}\right>\right> \equiv 
\Sigma_{\mathbf{k}\sigma}^{(U,sf)}(E)G_{\mathbf{k}\sigma}(E).
\end{equation}
$[..,..]_{-(+)}$ denotes the (anti)commutator. In general
$\Sigma_{\mathbf{k}\sigma}(E)$ will be a complex quantity. It contains 
all the influences of the various interaction processes being therefore
of fundamental importance. To have found $\Sigma_{\mathbf{k}\sigma}(E)$
means to have solved the problem. 

Another decisive quantity is the spectral density,
\begin{equation}
\label{eq:specdensity}
S_{\mathbf{k}\sigma}(E) = -\frac{1}{\pi}\Im G_{\mathbf{k}\sigma}(E),
\end{equation}
which is directly related to the bare lineshape of an angle- and
spin-resolved (inverse) photoemission experiment. Additional angle
averaging then leads to the quasiparticle density of states: 
\begin{equation}
\label{eq:qdos}
\rho_{\sigma}(E) =
\frac{1}{N\hbar}\sum_{\mathbf{k}}S_{\mathbf{k}\sigma}(E-\mu). 
\end{equation} 
We are going to present our results for the CKLM mainly in terms of
spectral densities and quasiparticle densities of states.

\subsection{Effective medium approach}
\label{subsec2}
Our study is predominantly focussed on the influence of the interband
exchange~(\ref{eq:Hsf2ndquant}) on the conduction band states. As 
explained in Sec. \ref{sec1} the introduction of $H_{U}$ shall first of
all avoid unphysical double occupancies as 
energetically favoured configurations. For this purpose we assume:
\begin{equation}
\label{eq:UWJ}
U \gg W, J.    
\end{equation}
$W$ is the bandwidth. In this limit the energy band is split for $J$=0
into two quasiparticle subbands. The $(\mathbf{k},\sigma)$-dependent
spectral density consists of two peaks 
centred at\cite{HAL60,HEN97A,BDN01}
\begin{eqnarray}
\label{eq:peakpos1}
\hspace{-4ex}&T_{1\sigma}(\mathbf{k})& = T_{0} +
(1-n_{-\sigma})(\epsilon(\mathbf{k})-T_{0})+n_{-\sigma}B_{-\sigma},\\[2ex]\label{peakpos2}
\hspace{-4ex}&T_{2\sigma}(\mathbf{k})& = T_{0}+U +
n_{-\sigma}(\epsilon(\mathbf{k})-T_{0})+(1-n_{-\sigma})B_{-\sigma}. 
\end{eqnarray}
To the same order, the area under the peaks (\textit{spectral weight})
is given by: 
\begin{equation}
\label{eq:specweights}
\hat{\alpha}_{1\sigma}(\mathbf{k}) =
1-n_{-\sigma}=1-\hat{\alpha}_{2\sigma}(\mathbf{k}). 
\end{equation}
$B_{-\sigma}$ is a higher correlation function, which, however, can
exactly be expressed by the spectral density itself:\cite{BDN01}   
\begin{align}
\label{eq:bshift}
B_{\sigma}&=\frac{1}{n_{\sigma}(1-n_{\sigma})}\frac{1}{N}\sum_{ij}^{i\neq j}T_{ij}\left<c_{i\sigma}
^{\dagger}c_{j\sigma}(2n_{i-\sigma}-1)\right>\nonumber\\
&=\frac{1}{n_{\sigma}(1-n_{\sigma})}\frac{1}{N}\sum_{\mathbf{k}}\left[\epsilon(\mathbf{k})-T_{0}\right]\nonumber\\
&\quad\times \int_{-\infty}^{+\infty}dE
f_{-}(E)S^{*}_{\mathbf{k}\sigma}(E-\mu)\left\{\frac{2}{U}\left[E-\epsilon(\mathbf{k})\right]-1\right\}.
\end{align}
$S^{*}_{\mathbf{k}\sigma}(E)$ is, at first, the spectral density for the
\textit{pure} Hubbard model; however, it will later be changed into the
spectral density of the full model to guarantee self-consistency.

In order to account for the Hubbard part $H_{U}$ in the strong coupling
regime we perform an effective medium approach by replacing the
single-particle energies in Eq.~(\ref{eq:Hs}) by the quasiparticle
energies~(\ref{eq:peakpos1}) and~(\ref{peakpos2}), respectively. This
means for the Green function~(\ref{eq:GformalS}): 
\begin{equation}
\label{eq:Gtot}
G_{\mathbf{k}\sigma}(E) = \sum_{i=1}^{2}G_{\mathbf{k}\sigma}^{(i)}(E),
\end{equation}
\vspace{-0.5cm}
\begin{equation}
\label{eq:Gparts}
G_{\mathbf{k}\sigma}^{(i)}(E)=\hbar\frac{\hat{\alpha}_{i\sigma}(\mathbf{k})}{E+\mu-T_{i\sigma}(\mathbf{k})-\Sigma_{\mathbf{k}\sigma}^{(i)}(E)}.
\end{equation}
The CKLM has therefore become a \textit{normal} KLM in an effective
medium, which is characterized by two (!) energy bands according to the
single-particle energies $T_{1,2\sigma}(\mathbf{k})$. The partial
self-energies $\Sigma_{\mathbf{k}\sigma}^{(1,2)}(E)$ are formally
determined by the interband-exchange, only.  

We are left with the many-body problem of the KLM which, for the general
case, cannot be treated rigorously either. However, some important facts
are known, which can help to find a trustworthy approach. 

\subsection{Zero-bandwidth limit}
\label{subsec3}
For a self-energy approach that is credible in the strong coupling
region ($JS$$\gg$$W$) the exactly calculable zero-bandwidth
limit\cite{NOM84} should be fulfilled: 
\begin{equation}
\label{eq:zeroW}
T_{ij} \rightarrow T_{0}\delta_{ij}~~;~~\epsilon(\mathbf{k}) \rightarrow
T_{0}~\forall~\mathbf{k}. 
\end{equation}
The \textit{free} conduction band is shrunk to an $N$-fold degenerate
level $T_{0}$. This means for the effective medium:
\begin{equation}
\label{eq:zeroWHubbard}
T_{1\sigma}(\mathbf{k}) \rightarrow T_{0}~~;~~T_{2\sigma}(\mathbf{k})
\rightarrow T_{0}+U. 
\end{equation}
To have a physically reasonable limit, however, the localized spin
system is furtheron considered as collectively ordered for temperatures
$T$$<$$T_{c}$ by any kind of exchange interaction. The respective
magnetization $\left<S^{z}\right>$ has to be considered as an external
parameter. The partial self-energies as defined in (\ref{eq:Gparts})
read\cite{NOM84} 
\begin{align}
\label{eq:zeroWsigma1}
\Sigma_{\sigma}^{(1)(W=0)}(E)&=\frac{1}{2}J\frac{\frac{1}{2}JS(S+1)-X_{-\sigma}(E+\mu-T_{0})}
{E+\mu-T_{0}-\frac{1}{2}J(1+X_{-\sigma})},\\
\label{eq:zeroWsigma2}
\Sigma_{\sigma}^{(2)(W=0)}(E)&=\frac{1}{2}J\frac{\frac{1}{2}JS(S+1)+Y_{-\sigma}(E+\mu-T_{0}-U)}
{E+\mu-T_{0}-U+\frac{1}{2}J(1+Y_{-\sigma})}.
\end{align}
Here we have introduced:
\begin{equation}
\label{eq:abbrevXY}
X_{\sigma}=\frac{\Delta_{\sigma}-m_{\sigma}}{1-n_{\sigma}}~~~,~~~
Y_{\sigma}=\frac{\Delta_{\sigma}}{n_{\sigma}}~~~,
\end{equation}
where
\begin{equation}
\label{eq:abbrevmDelta}
m_{\sigma}=z_{\sigma}\left<S^{z}\right>~~~,~~~
\Delta_{\sigma}=\left<S^{\sigma}_{i}c^{\dagger}_{i-\sigma}c_{i\sigma}\right>
+ z_{\sigma}\left<S^{z}_{i}n_{i\sigma}\right>. 
\end{equation}

\subsection{Magnetic polaron}
\label{subsec4}
There is another very instructive limiting case that can be solved
rigorously.\cite{SM81,NDM85} It concerns a single electron (hole) in an
otherwise empty (fully occupied) energy band interacting with a
ferromagnetically saturated spin system. The details of the derivation
of the self-energy can be found in Ref.~\onlinecite{NOL702}. If we
introduce the \textit{effective medium propagators} ($i$=1,2) 
\begin{equation}
\label{eq:effmedprop}
G_{i\sigma}(E)=\frac{1}{N}\sum_{\mathbf{k}}\frac{\hbar}{E+\mu-T_{i\sigma}(\mathbf{k})}
\end{equation}
then we have the following $T$=0-expressions for the self-energy, which
are exact for arbitrary bandwidth, coupling, and spin:
\begin{align}
&\Sigma_{\mathbf{k}\sigma}^{(n=0)}(E) \equiv \Sigma_{\sigma}^{(1)(n=0)}(E)\nonumber\\
\label{eq:sigmapolaron1}
&~=-\frac{1}{2}z_{\sigma}JS + \frac{1}{4}J^{2}
\frac{(1-z_{\sigma})SG_{1-\sigma}(E-\frac{1}{2}z_{\sigma}JS)} 
{1-\frac{1}{2}JG_{1-\sigma}(E-\frac{1}{2}z_{\sigma}JS)},
\end{align}
\begin{align}
&\Sigma_{\mathbf{k}\sigma}^{(n=2)}(E) \equiv \Sigma_{\sigma}^{(2)(n=2)}(E)\nonumber\\
\label{eq:sigmapolaron2}
&~=-\frac{1}{2}z_{\sigma}JS + \frac{1}{4}J^{2}
\frac{(1+z_{\sigma})SG_{2-\sigma}(E-\frac{1}{2}z_{\sigma}JS)} 
{1+\frac{1}{2}JG_{2-\sigma}(E-\frac{1}{2}z_{\sigma}JS)}.
\end{align}

\subsection{Weak-coupling behaviour}
\label{subsec5}
Since conventional diagrammatic perturbation theory for the KLM is
impossible due to the lack of Wick's theorem we determine the
weak-coupling behaviour by use of the projection-operator
method.\cite{MOR65,MOR66} Strictly applying the Mori-formalism to the
effective-medium KLM yields up to terms $J^{2}$:\,\cite{HIC01} 
\begin{align}
&\Sigma_{\mathbf{k}\sigma}^{s.o.}(E)=-\frac{1}{2}Jm_{\sigma}-\frac{1}{4}J^{2}m^{2}_{\sigma}G^{(0)}_{\mathbf{k}\sigma}(E)\nonumber\\
&+\frac{J^2}{4N^{2}}\sum_{\mathbf{q}}\bigg[\left<S_{-\mathbf{q}}^{z}S_{\mathbf{q}}^{z}\right>
G_{\mathbf{k}+\mathbf{q}\sigma}^{(0)}(E)\bigg.\nonumber\\
\label{eq:sigma2ndorder}
&~~~~+\bigg.\left(\left<S_{-\mathbf{q}}^{-\sigma}S_{\mathbf{q}}^{\sigma}\right> 
+2z_{\sigma}\left<S_{\mathbf{0}}^{z}\right>\left<n_{\mathbf{k}+\mathbf{q}-\sigma}\right>^{(0)}\right)
G_{\mathbf{k}+\mathbf{q}-\sigma}^{(0)}(E)\bigg].
\end{align}
The $\mathbf{q}$-dependent spin operator is defined as usual:
\begin{equation}
\label{eq:Sq}
S_{\mathbf{q}}^{\alpha}=\sum_{i}S_{i}^{\alpha}e^{-i\mathbf{q}\mathbf{R}_{i}},    
\end{equation}
($\alpha$=+,-,z). $G_{\mathbf{k}\sigma}^{(0)}(E)$ is the
effective medium Green function~(\ref{eq:Gtot}) in case of a vanishing
self-energy. In the following we are mainly interested in the local
self-energy,  
\begin{equation}
\label{eq:sigmalocal}
\Sigma_{\sigma}(E)=\frac{1}{N}\sum_{\mathbf{k}}\Sigma_{\mathbf{k}\sigma}(E),
\end{equation}
for which the self-energy (\ref{eq:sigma2ndorder}) further simplifies.

\subsection{High-energy expansion}
\label{subsec6}
The spectral moments $M_{\mathbf{k}\sigma}^{(n)}$ of the spectral
density~(\ref{eq:specdensity}), 
\begin{equation}
\label{eq:specmoments1}
M_{\mathbf{k}\sigma}^{(n)}=\frac{1}{\hbar}\int_{-\infty}^{+\infty}dEE^{n}S_{\mathbf{k}\sigma}(E),
\end{equation}
can be of great importance for testing or constructing unavoidable 
approximations. This is due to the fact that in principle, the moments
can be calculated rigorously and independently of the required spectral
density: 
\begin{align}
\label{eq:specmoments2}
M_{\mathbf{k}\sigma}^{(n)}=\left[\left(i\hbar\frac{\partial}{\partial
t}\right)^{n}\left<\left[c_{\mathbf{k}\sigma}(t),
c_{\mathbf{k}\sigma}^{\dagger}(t')\right]_{+}\right>\right]_{t=t'},\nonumber\\
(n=0,1,2,..).
\end{align}
There is a close connection between the moments and the high-energy
behaviour of the Green function: 
\begin{equation}
\label{eq:highEexp}
G_{\mathbf{k}\sigma}(E)=\int_{-\infty}^{+\infty}\frac{S_{\mathbf{k}\sigma}(E')}{E-E'}
=\hbar\sum^{\infty}_{n=0}\frac{M_{\mathbf{k}\sigma}^{(n)}}{E^{n+1}}.
\end{equation}
This transfers to the self-energy via the Dyson equation:
\begin{equation}
\label{eq:highEsigma}
\Sigma_{\mathbf{k}\sigma}(E)=\sum^{\infty}_{m=0}\frac{C_{\mathbf{k}\sigma}^{(m)}}{E^{m}}.
\end{equation}
The coefficients $C_{\mathbf{k}\sigma}^{(m)}$ are simple functions of
the moments up to order $m$+1.\cite{NRRM01} In 
the limits $n$$\rightarrow$0 and $n$$\rightarrow$2, respectively, we can
write down corresponding formulas for the partial self-energies in
Eq.~(\ref{eq:Gparts}), which we use in the next section to fix 
free parameters in our basic self-energy ansatz.

\section{Interpolating self-energy ansatz}
\label{sec3}
We want to develop a self-energy approach which fulfills a maximum
number of exactly known limiting cases. So the zero-bandwidth
case~(\ref{eq:zeroWsigma1}), (\ref{eq:zeroWsigma2}) should correctly be  
reproduced for all temperatures $T$, band occupations $n$, and coupling
strengths $J$. Furthermore, the non-trivial
results~(\ref{eq:sigmapolaron1}), (\ref{eq:sigmapolaron2}) for
ferromagnetic saturation, valid for arbitrary bandwidths and couplings,
are strong criteria for the self-energy approach. In addition we have
the result~(\ref{eq:sigma2ndorder}) for the weak-coupling regime holding
for all bandwidths and temperatures. All these exact facts can be
covered by the following structures of the partial self-energies:
\begin{align}
\label{eq:sigmaansatz1}
\Sigma_{\sigma}^{(1)}(E)&=-\frac{1}{2}JX_{-\sigma}+
\frac{1}{4}J^{2}\frac{a_{-\sigma}G_{1-\sigma}(E-\frac{1}{2}JX_{-\sigma})}
{1-b_{-\sigma}G_{1-\sigma}(E-\frac{1}{2}JX_{-\sigma})},\\
\label{eq:sigmaansatz2}
\Sigma_{\sigma}^{(2)}(E)&=\frac{1}{2}JY_{-\sigma}+
\frac{1}{4}J^{2}\frac{\hat{a}_{-\sigma}G_{2-\sigma}(E+\frac{1}{2}JY_{-\sigma})}
{1+\hat{b}_{-\sigma}G_{2-\sigma}(E+\frac{1}{2}JY_{-\sigma})}.
\end{align}
Here we have assumed for simplicity a local,
i.e. \mbox{$\mathbf{k}$-independent} self-energy, as is the case for all
our above-listed exact limiting cases. This assumption is not necessary,
but makes some steps of the evaluation easier and a bit more 
transparent. $X_{\sigma}$ and $Y_{\sigma}$ are defined
in~(\ref{eq:abbrevXY}), the effective medium propagators $G_{i\sigma}(E)$
in~(\ref{eq:effmedprop}). 

Equations~(\ref{eq:sigmaansatz1}) and~(\ref{eq:sigmaansatz2}) contain 
not yet fixed parameters $a_{\sigma},~ b_{\sigma},~\hat{a}_{\sigma}$,
and $\hat{b}_{\sigma}$. They can be derived by use of the high-energy
expression~(\ref{eq:highEsigma}). In order to do so for the general
case, we have to determine four unknown parameters, requiring the 
first five terms in the expansion~(\ref{eq:highEsigma}). On the other
hand, for the determination of $C_{\mathbf{k}\sigma}^{(m)}$ for
$m$=0,1,..,4, the derivation of the first six spectral moments is
needed. This turns out to be impossible, in particular because higher 
(mixed) correlation functions appear in the moments of order $n$$\geq$3,
which cannot be determined self-consistently. However, simplifications
show up in the limits $n$$\rightarrow$0 and $n$$\rightarrow$2, for which
we can separately inspect the partial self-energies needing only the
first four moments for, respectively, the empty and the occupied band. 

Because of the local character of the partial self-energies
$\Sigma_{\sigma}^{(1,2)}(E)$ we need the local self-energy coefficients:
\begin{equation}
\label{eq:selfElocalcoeff}
C_{\sigma}^{(m)}=\frac{1}{N}\sum_{\mathbf{k}}C_{\mathbf{k}\sigma}^{(m)}.
\end{equation}
A tedious but straightforward calculation eventually yields for the
unknown parameters in~(\ref{eq:sigmaansatz1}) and
(\ref{eq:sigmaansatz2}): 
\begin{align}
\label{eq:a}
a_{\sigma}&=S(S+1)-X_{\sigma}(X_{\sigma}+1),\\
\label{eq:ahat}
 \hat{a}_{\sigma}&=S(S+1)-Y_{\sigma}(Y_{\sigma}+1),\\
\label{eq:b}
b_{\sigma}&=\hat{b}_{\sigma}=\frac{1}{2}J.
\end{align}
It is easy to check that all the rigorous limiting cases we listed up 
before, are strictly fulfilled by the now complete result for the
self-energy. Eqs.~(\ref{eq:peakpos1}) to (\ref{eq:Gparts}), 
(\ref{eq:abbrevXY}), (\ref{eq:abbrevmDelta}),
(\ref{eq:sigmaansatz1}), (\ref{eq:sigmaansatz2}), and~(\ref{eq:a}) to
(\ref{eq:b}) build a close system of equations that can be solved
self-consistently for the CKLM as soon as we can express the  
correlation functions $n_{\sigma}$ and $\Delta_{\sigma}$ by the
single-electron Green function~(\ref{eq:Gtot}). This is 
no problem for $n_{\sigma}$ because we can use the spectral theorem to 
get: 
\begin{equation}
\label{eq:n_sigma}
n_{\sigma}=-\frac{1}{\pi N}\sum_{\mathbf{k}}\sum_{i=1}^{2}\int_{-\infty}^{+\infty}dEf_{-}(E)\Im
G_{\mathbf{k}\sigma}^{(i)}(E-\mu),
\end{equation}
where $f_{-}(E)=\left(1+e^{\beta(E-\mu)}\right)^{-1}$ denotes the Fermi
function. Even the mixed correlation function $\Delta_{\sigma}$, defined
in~(\ref{eq:abbrevmDelta}), can rigorously be expressed by
$G_{\mathbf{k}\sigma}(E)$. Using the spectral theorem for higher Green
functions, properly defined with respect to the two terms
of $\Delta_{\sigma}$ in~(\ref{eq:abbrevmDelta}), and exploiting the
equation of motion of $G_{\mathbf{k}\sigma}(E)$, one gets: 
\begin{align}
\Delta_{\sigma}&=\frac{2}{NJ\pi}\sum_{\mathbf{k}}\sum_{i=1}^{2}
\int_{-\infty}^{+\infty}dEf_{-}(E)\left[E-T_{i\sigma}(\mathbf{k})\right]\nonumber\\
\label{eq:Delta_sigma}
&\hspace{4cm}\times\Im G_{\mathbf{k}\sigma}^{(i)}(E-\mu).
\end{align}
The local moment magnetization $\left<S^{z}\right>$ shall be considered
as an external parameter being responsible for the induced
temperature-dependence of the band states. 

Once again we want to stress that all the calculations have been
done for quantum mechanical spins, which is in contrast to many of the 
recent works on the KLM.\cite{FUR99,MLS95,HEV00} 

\section{Results}
\label{sec4}
For the presentation of the results of our theory on the CKLM we have
chosen an sc-lattice with the respective Bloch density of states (BDOS)
in tight-binding approximation.\cite{JEL69} For all calculations we have
assumed a bandwidth of 1 eV while the center of gravity of the
\textit{free} Bloch band defines the energy zero. For all 
evaluations we have fixed the Coulomb interaction to $U$=2~eV. 

\begin{figure}[t]
\includegraphics[width=0.45\textwidth]{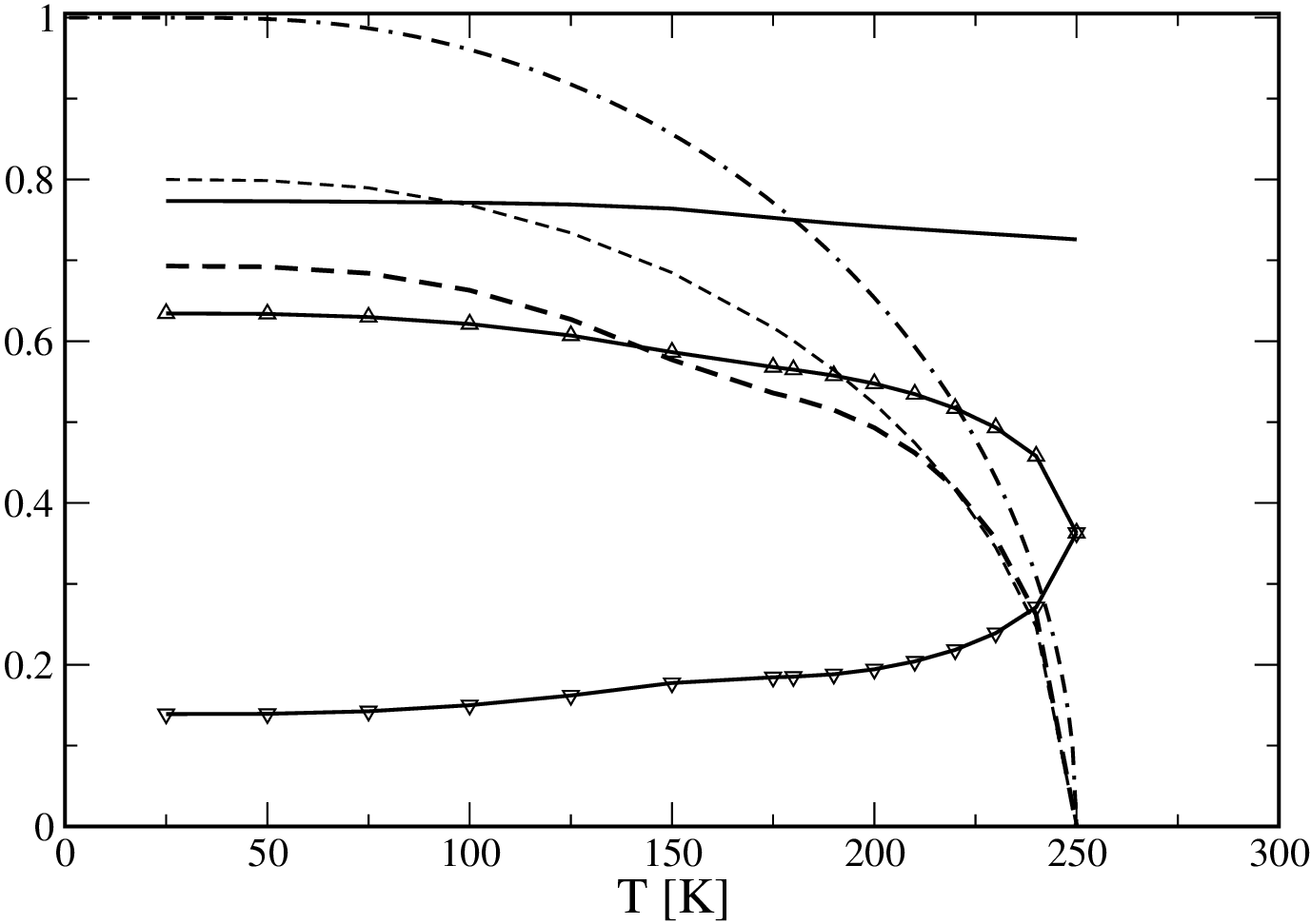}
\caption{Temperature dependence of the correlation
  functions. Dash-dotted line $\left<S^{z}\right>/S$, thin dashed line
  renormalized $\left<S^{z}\right>$ [see Eq.~(\ref{eq:Sz_renorm}) and text],
  thick dashed line electron spin-polarization [see
  Eq.~(\ref{eq:el_polarization})], $\Delta$-line~$\Delta_\uparrow$,
  $\nabla$-line~$\Delta_\downarrow$, full line
  $\Delta_\uparrow+\Delta_\downarrow$. Parameters:
  $S=~3/2,~n=~0.5,~U=2$~eV, $J=1$~eV, $T_{c}=250$~K.}
\label{corr_functions}
\end{figure}
\begin{figure}[t]
\includegraphics[width=0.48\textwidth]{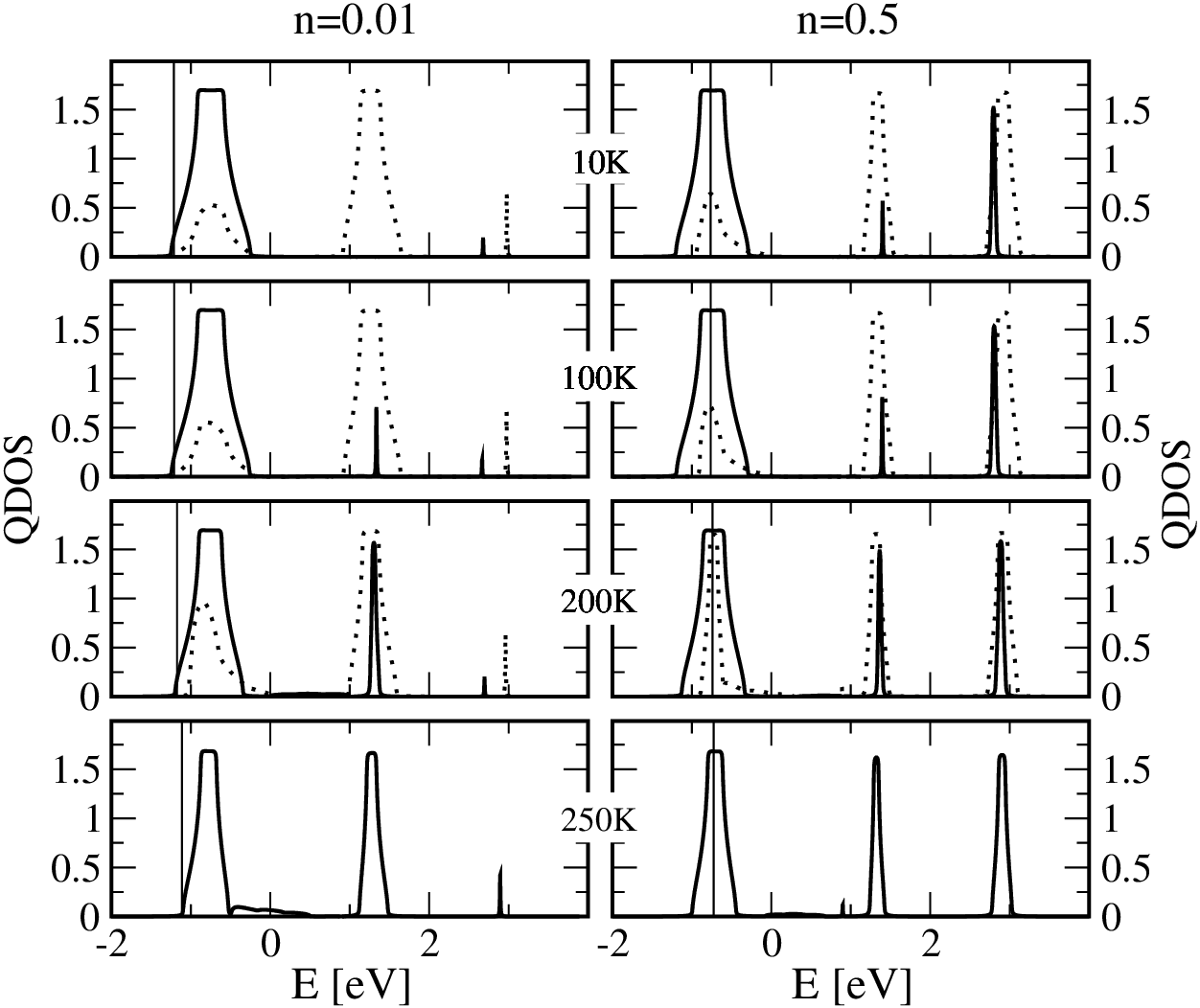}
\caption{Quasiparticle density of states for various values
  of the temperature. Left low-density case, right intermediate band
  occupation. Full line $\uparrow$ and dotted line $\downarrow$.
  Parameters: $S=3/2,~U=2$ eV, $J=1$~eV. The chemical
  potential is indicated by a thin vertical line.}
\label{qdos_Tscan}
\end{figure}
Our theory does not aim at a self-consistent determination of the 
local-moment magnetization \cite{SN02,NRM97} but rather at the influence
of the interband exchange on the band states. We therefore consider
$\left<S^{z}\right>$ as a parameter, for which we have chosen a
Brillouin function according to the spin value $S$=3/2 with a
Curie temperature of $T_{c}$=250~K (Fig.~\ref{corr_functions}). One has,
however, to bear in mind that not all parameter constellations will
permit a finite magnetization. Furthermore, saturation
$\left<S^{z}\right>$=$S$ may not be reachable by the system in case of
finite band occupation due to spin exchange processes. This can be
demonstrated directly by the exact zero-bandwidth solution.\cite{NOM84}
For in this special case results get unphysical for less than
half-filled bands ($n$$\leq$1) as soon as  
\begin{equation}
\label{eq:Sz_renorm}
\left<S^{z}\right>\geq S\cdot\frac{S+1-n}{S+1}. 
\end{equation}
We observe in our theory a similar effect: results may become unphysical
when for finite band occupations $\left<S^{z}\right>$ exceeds a critical
value which depends on $n$ as well as on $J$/$W$. For decreasing $J$/$W$
the \textit{demagnetization factor} becomes smaller, disappearing for
$J$/$W$$\rightarrow$0. To bring a certain systematics to our results we
have therefore renormalized the parameter \textit{magnetization}
$\left<S^{z}\right>$ by the factor $\frac{S+1-n}{S+1}$. If we speak in
the following of temperature then this is thought to be connected to a
magnetization given by a Brillouin function multiplied by this
factor. As an example we have plotted in Fig.~\ref{corr_functions} the
renormalized magnetization for $n$=0.5 (thin dashed line).  

Figure~\ref{qdos_Tscan} (left) exhibits the temperature-dependence of the QDOS
of the CKLM for the case of a very low band occupation
$n$=0.01. $J$=1~eV belongs already to the strong coupling region. The 
QDOS mainly consists of two subbands for each spin direction. Each of
these subbands has a clear physical meaning. Roughly speaking, in the
lower (upper) band the electron has oriented its spin parallel
(antiparallel) to the localized spin. The distance of the centers of
gravity of the subbands is close to $\frac{1}{2}J$(2$S$+1) ($\approx$
2~eV), corresponding to the distance of the two respective energy levels
in the zero-bandwidth limit.\cite{NOM84} $T$=10~K means that the spin system is
almost ferromagnetically saturated. This is a special case, for which
our theory turns out to be exact (Sec.~\ref{subsec4}). The
$\uparrow$-electron has no chance to exchange its spin with the parallel
aligned localized spin system. From 
the exchange interaction~(\ref{eq:Hsf2ndquant}) only the Ising-like part
works, simply leading to a rigid shift of the total spectrum. The
$\uparrow$-QDOS is identical to the sc-BDOS except for the rigid
shift. On the other hand, the $\downarrow$-electron has two
possibilities for a spinflip. It can emit a magnon therewith becoming
itself an $\uparrow$-particle. Such a process can happen only if there
are $\uparrow$-states within reach that the excited
$\downarrow$-electron may occupy after the spinflip caused by the magnon
emission. This is the reason why the lower $\downarrow$-subband, which
consists of such \textit{scattering states}, covers exactly the same
energy region as $\rho_{\uparrow}(E)$. The $\downarrow$-electron has a
second possibility to exchange its spin. It can polarize its nearest
spin neighborhood by repeated magnon emission and reabsorption, thus
propagating through the lattice as a dressed particle with a virtual
cloud of magnons. We call it then the \textit{magnetic polaron}. It has
its analogue in the Fr\"{o}hlich polaron of lattice dynamics. Such
polaron states build up the second subband.  
\begin{figure}[b]
\includegraphics[width=0.48\textwidth]{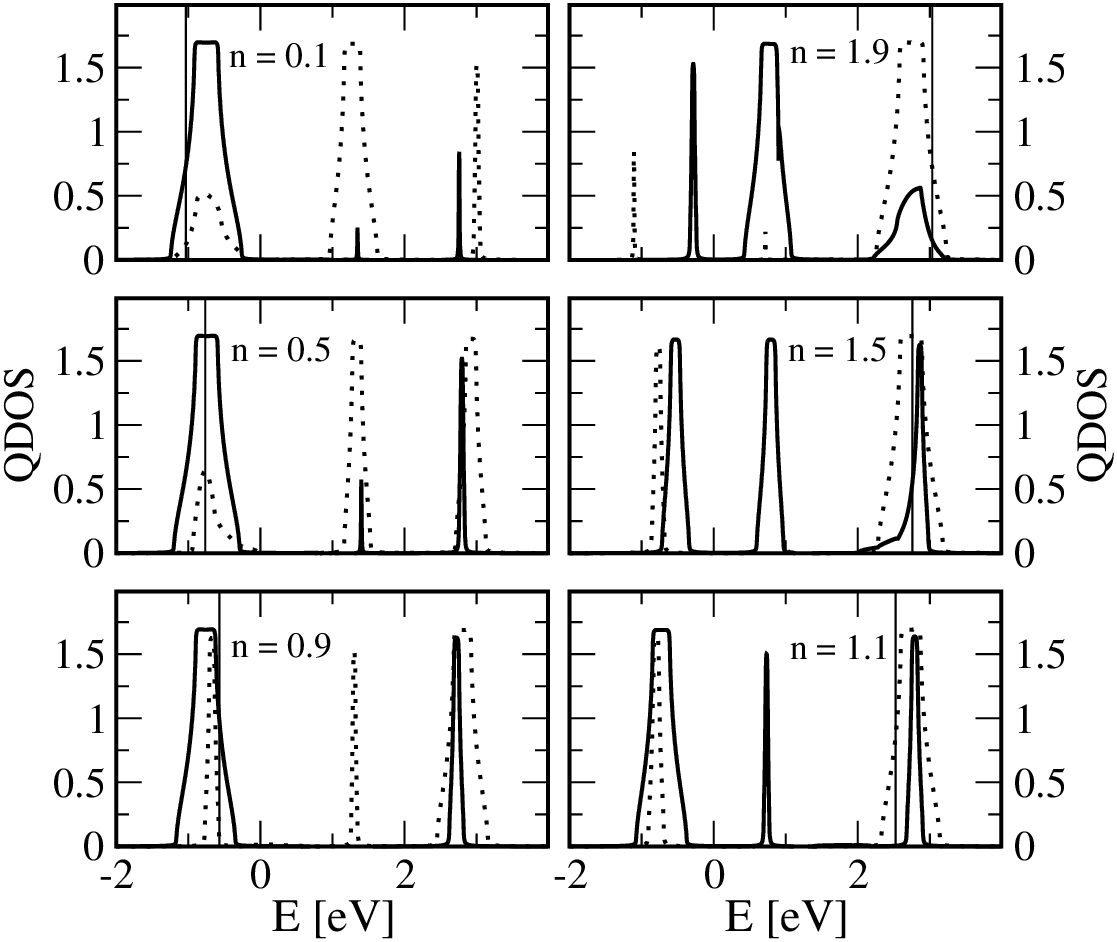}
\caption{Quasiparticle density of states for various values of the band
  occupation (ferromagnetic region). Full line for spin up and dotted
  line for spin down. Parameters: $S=3/2,~U=2$ eV, $J=1$ eV, and
  $T=10$ K.}    
\label{qdos_nscan_T10}
\end{figure}

For increasing temperature $T$ (i.e. for decreasing magnetization), the
$\uparrow$-spectrum, too, becomes more complicated. Magnon absorption by
a $\uparrow$-electron is equivalent to magnon emission by the
$\downarrow$-electron, with the only exception that emission is always
possible, absorption, however, only if there are magnons available. This
is not the case in ferromagnetic saturation. Consequently the
$\uparrow$-spectrum is then relatively simple. With increasing
temperature, however, the two spin spectra become more and more similar,
until at $T_{c}$ the spin asymmetry has disappeared. The quasiparticle
splitting remains, though.

Let us now discuss the dependence on the band occupation. In
Fig.~\ref{qdos_nscan_T10} the QDOS is plotted for various band
occupations at $T$=10~K, i.e., for practically maximum local spin
magnetization. First of all, we observe the appearance of a
third quasiparticle subband due to possible double occupancies of
lattice sites, which require the additional Coulomb energy
$U$. According to the exact zero-bandwidth limit one 
could expect for less than half-filled bands even a second high-energy
subband.\cite{NOM84} This is not the case because one of the upper bands
has a vanishing spectral weight. This fact already holds in the
$W$=0-case. For more than half-filled bands one of the low-energy 
subbands disappears.

Two details of the occupation dependence of the QDOS are worth to be
stressed. First we learn from the \mbox{$n$-dependent} position of the chemical
potential in the ferromagnetic phase (Fig.~\ref{qdos_nscan_T10}) that
the electron system is never completely spin-polarized. This is a
natural consequence of the spin exchange processes between itinerant and
localized electrons and in contradiction to several other treatments of
the KLM.\cite{HEV00,MMS96} Total spin polarization is only true under
the unphysical assumption of classical spins. The second point concerns 
the polaron band, which for less than half-fillings is never occupied 
(Fig.~\ref{qdos_nscan_T10}). The chemical potential
is always in the lowest subband reaching its upper edge for
$n$$\rightarrow$1. At the same moment the polaron band disappears. The
chemical potential jumps into the uppermost subband for more than
half-fillings guaranteeing therewith particle-hole symmetry. Note that
the middle band in case of $n$$>$1 is the polaron band for holes, centred
at $U$$-$$\frac{1}{2}J$($S$+1).

\begin{figure}[b]
\includegraphics[width=0.4\textwidth]{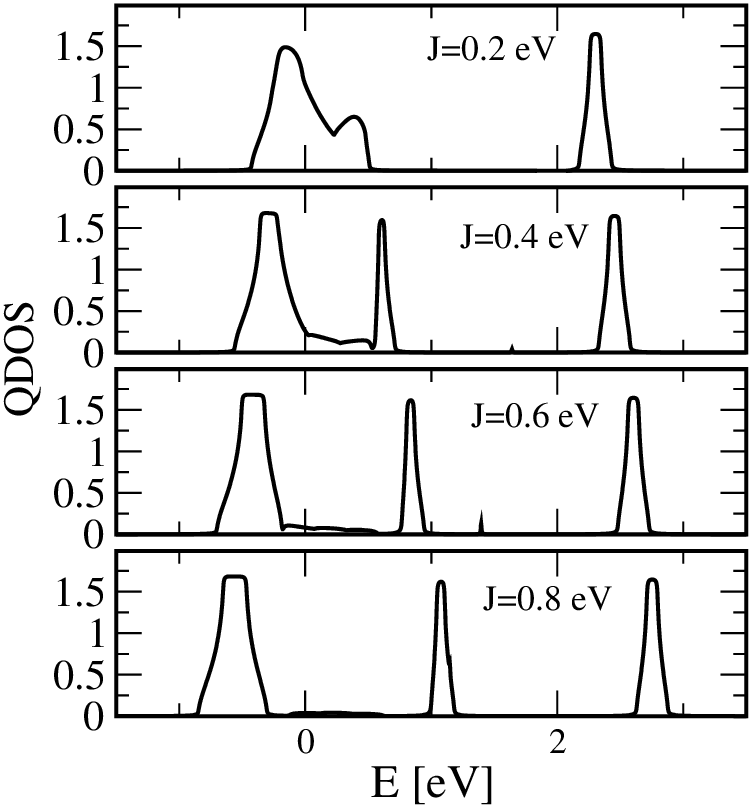}
\caption{Quasiparticle density of states for various values of the Hund 
  coupling $J$. Parameters: $S~=~3/2,~U~=~2$~eV, $n=0.5$, and $T=250$ K.}
\label{qdos_Jscan_Tc}
\end{figure}

Figure~\ref{qdos_Jscan_Tc} demonstrates the $J$-dependence of the
(paramagnetic) QDOS for a band filling of $n$=0.5. For weak (moderate)
couplings the two lower subbands overlap, separating, however, as soon
as $J$ exceeds a critical value. The existence of the two low-energy
subbands can roughly be understood from the zero-bandwidth limit. In the
strong coupling regime ($JS$$\gg$$W$) the 
band electron has only a low mobility being trapped by the local spin
for a rather long time. In such a case the local spin defines the
quantization axis. When we speak of \textit{up spin band} or
\textit{down spin band} we refer the electron spin to an external
axis. However, in the strong coupling regime the relative orientation of
the electron spin in the local frame is decisive for the excitation
energies. The lowest subands, up-spin as well as down-spin,  consists of
states where the excited electron enters directly or after magnon
emission (absorption) the local frame parallel to the local spin. If it
arrives at the local frame antiparallel to the local spin then it goes
into the second low-energy subband centered at
$+\frac{1}{2}J$($S$+1). There is a slight probability that the electron is
not trapped by the localized spin but rather propagating with high
mobility through the spin lattice. Then the total magnetization
$\left<S^{z}\right>$ will become decisive as effective quantization
axis, leading to a mean-field type behaviour of the quasiparticle
according to the first term in the
self-energy~(\ref{eq:sigmaansatz1}). We interpret the small shoulder in
between the two lower subbands in Fig.~\ref{qdos_Jscan_Tc} in such a
way as already discussed in an earlier work.\cite{NRRM01} This shoulder
becomes more and more unimportant with increasing exchange coupling.

Let us now inspect the temperature dependence of the quasiparticle
spectrum for a higher band occupation. Figure~\ref{qdos_Tscan} (right)
shows as an example the QDOS for $J$=1~eV and $n$=0.5. The positions of
the bands are hardly affected by temperature as one might expect because
of the first term in the self-energy~(\ref{eq:sigmaansatz1}) and
(\ref{eq:sigmaansatz2}), respectively. For strong exchange couplings the
second term, which incorporates all the spinflip processes,
dominates. As explained just before, the local frame is decisive being
of course unaffected by temperature. The temperature determines,
however, the probability with which an externally prepared
$\sigma$-$\textit{electron}$ enters the local frame as 
$\uparrow$-($\downarrow$-)\textit{electron}, and this fixes the spectral 
weights of (areas under) the subband-DOS. For example, in the
ferromagnetically saturated spin system ($T$=0) an $\uparrow$-electron
cannot enter the local frame antiparallel to the localized
spin. Therefore the middle subband disappears because of vanishing 
spectral weight. This changes with increasing temperature,
i.e. decreasing magnetization by directional disorder of the local 
spins. For $T$$\geq$$T_{c}$ the spin asymmetry has disappeared, but
there remains a quasiparticle splitting into three peaks, which are all
of comparable spectral weights. For strongly coupled Kondo-lattice
systems such an unconventional splitting should be observable in a
respective photoemission experiment. 

\begin{figure}[t]
\includegraphics[width=0.45\textwidth]{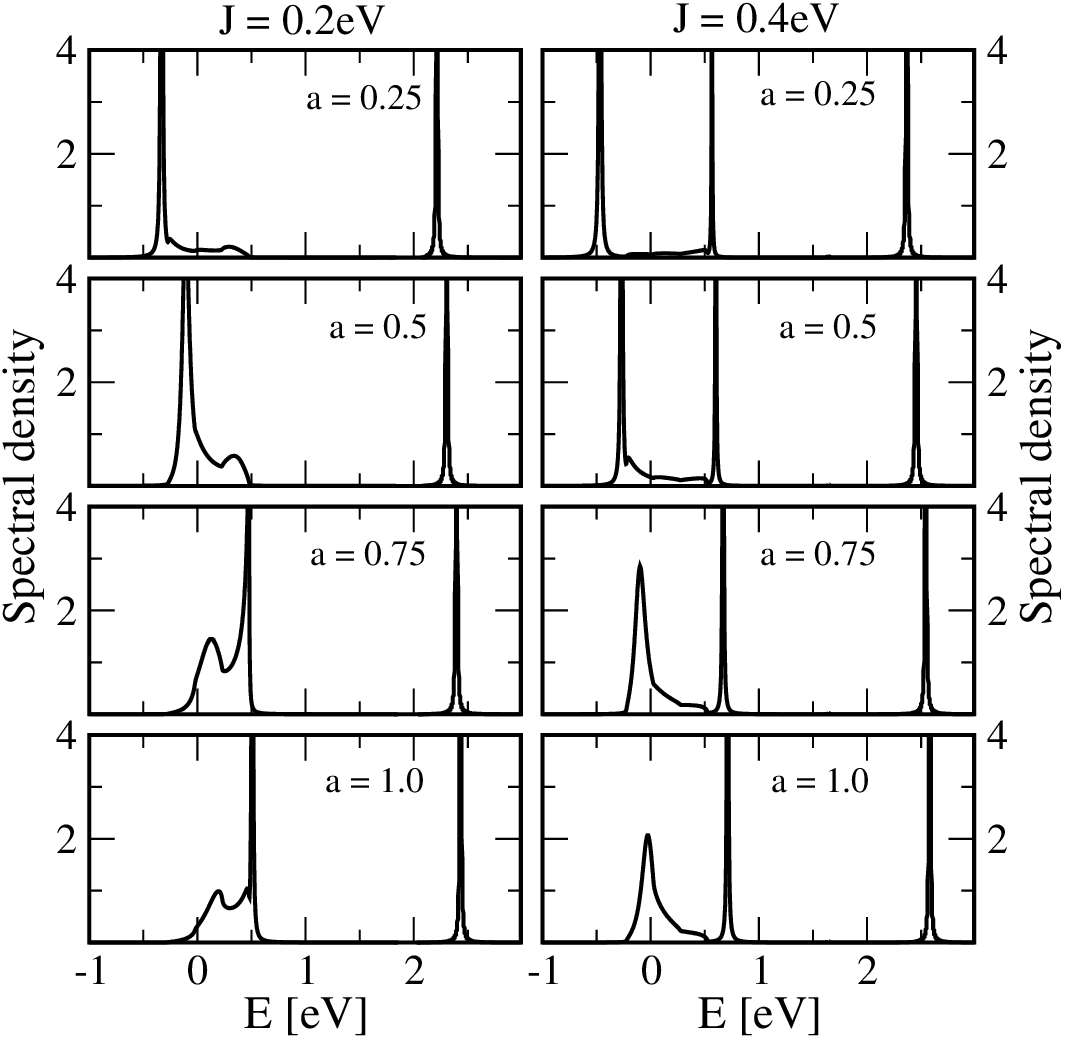}
\caption{Spectral density for two values of $J$ and three values of
  ${\bf k}$ along the ($a,a,a$)-direction. Parameters: $S~=~3/2,~U=2$ eV,
  $n=0.5$, and $T=250$ K.}  
\label{specdens3}
\end{figure}

Up to now we have only discussed the strong coupling case which we are
mainly interested in here. To give an example for weaker couplings we
present in Fig.~\ref{specdens3} the spectral density
(\ref{eq:specdensity}) for $T=T_{c}$ and two different exchange
couplings $J$. The low-energy part belongs to scattering states and the
magnetic polaron. The high-energy peak is due to double occupancies,
fairly uninteresting in the case of a less than half-filled band. For
$J$=0.2~eV the polaron peak dips into the scattering spectrum; the
quasiparticle therefore gets a finite lifetime. There is a remarkable
${\bf k}$-dependent shift of spectral weight. For a ${\bf k}$-vector
from the bottom or from the top of the respective quasiparticle subband
the magnetic polaron is rather stable, while experiencing substantial
damping in between. In systems with stronger exchange couplings
($J$=0.4~eV) the polaron peak, negligibly damped and hardly
dispersive, splits off the scattering part. It contributes to a very
narrow quasiparticle subband (see Fig.~\ref{qdos_Jscan_Tc}).

By use of Eqs.~(\ref{eq:n_sigma}) and (\ref{eq:Delta_sigma}) we can 
eventually evaluate the average spin-dependent particle density
$n_{\sigma}$ and the mixed correlation $\Delta_{\sigma}$, and therewith
the conduction band polarization, 
\begin{equation}
\label{eq:el_polarization}
m=\frac{n_{\uparrow}-n_{\downarrow}}{n_{\uparrow}+n_{\downarrow}}, 
\end{equation}
and the spin scalar product which appears in the exchange Hamiltonian 
(\ref{eq:Hsf}): 
\begin{equation}
\label{eq:spin_scalar}
\left<\mathbf{\sigma}_{i}\cdot\mathbf{S}_{i}\right>=\Delta_{\uparrow}+\Delta_{\downarrow}.
\end{equation}
A typical example ($J$=1~eV, $n$=0.5) is plotted in
Fig.~\ref{corr_functions}. We recognize that even for strong couplings 
the electron spin polarization is far from being complete. No more than
70\% polarization are observed. This again strongly contradicts other
treatments of the KLM.\cite{HEV00,MMS96} The higher correlation
functions $\Delta_{\uparrow}$ and $\Delta{\downarrow}$ exhibit a
distinct temperature-dependence, while the sum of both makes the
spin-spin scalar product~(\ref{eq:spin_scalar}) almost
temperature-independent.    

\section{Conclusions}
\label{sec5}
We have presented an interpolating self-energy approach to the many-body
problem of the \textit{correlated Kondo-lattice model} which
interpolates between a maximum number of rigorous statements and exact
limiting cases of the (in general) not exactly solvable
model. Correlations between the band electrons are accounted for by an
on-site Coulomb interaction of Hubbard type. In this paper we were not
interested in the investigation of how the band electron correlations
effect the magnetic stability within the KLM, as was done, for
instance, in Ref.~\onlinecite{HEV00}. A corresponding study of the
(self-consistent) magnetic part is currently being done and will be
presented at a later stage. Here the mentioned correlations simply
help to avoid unphysical double occupancies of lattice sites as
energetically favored excitations and to separate respective
quasiparticle subbands. 

In our theory the strongly exchange coupled system is accounted for by
the exact zero-bandwidth limit for all temperatures and band
occupations. The weakly coupled system is addressed by second order
perturbation theory (projection-operator technique) for all temperatures
and bandwidths. For further interpolation between the limits we used the
exact but non-trivial results for the special case of a single electron
(hole) in an otherwise empty (fully occupied) band interacting with a
ferromagnetically saturated spin system. This special case exhibits
practically all typical elementary processes that determine the physics
of the CKLM. It holds for all coupling strengths and bandwidths. Finally
we guaranteed the correctness of spectral moments via fitting free
parameters in the self-energy ansatz. 
 
The energy spectra show strong dependencies on temperature and band
occupation, particularly in the strongly exchange coupled regime which
is thought to be relevant for the \textit{colossal magnetoresistance
  materials} as well as for manganese-doped GaAs
(\textit{spintronics}). In our opinion the results clearly demonstrate
the importance to retain the quantum nature of the localized spins. To
replace them, for technical reasons, by classical spins appears to be
inadequate.  
 
It must be considered as a shortcoming of the approach that the local
moment magnetization had to come into play only as an external
parameter, which can give rise to certain inconsistencies. It is planned
for the future to incorporate the local moment system into the
self-consistency circle. 

\section*{Acknowledgments}
This work was prepared as an India-Germany Partnership Program sponsored
by the Volkswagen Foundation. Financial support by the Deutsche
Forschungsgemeinschaft within the Sonderforschungsbereich 290 is also
gratefully acknowledged as well as support by the Friedrich-Naumann
Foundation for one of us (J.K.).

\end{document}